\documentclass[a4,reqno]{article} 

 \usepackage{graphicx} \usepackage{amsmath}

\suppressfloats[t]

\begin{document} \large

\begin{center}
{\bf Numerical Simulation of Collinear Capillary-Wave Turbulence}\\[1.0ex]

\vspace{2mm}
{\it Evgeny Kochurin$^{1*}$, Guillaume Ricard$^2$,  Nikolay Zubarev$^{1,3}$, Eric Falcon$^2$ }\\[1.0ex]
{$^1$Institute of Electrophysics, Ural Division, Russian Academy of Sciences, 106 Amundsen Street, 620016 Ekaterinburg, Russia\\
$^2$Universit\'e de Paris, Univ. Paris Diderot, MSC Laboratory, UMR 7057 CNRS, F-75 013 Paris, France\\
$^3$P.N. Lebedev Physical Institute, Russian Academy of Sciences 119991, 53 Leninskij prospect, Moscow, Russia\\
*E-mail: kochurin@iep.uran.ru
}
\end{center}

\vspace{2mm}

\begin{abstract}
\large
We report on direct numerical simulation of quasi-one-dimensional bidirectional capillary-wave turbulence. Although nontrivial three-wave and four-wave resonant interactions are absent in this peculiar geometry, we show that an energy transfer between scales still occurs concentrated around the linear dispersion relation that is broadened by nonlinearity.  The wave spectrum displays a clear wave number power-law scaling that is found to be in good agreement with the dimensionally prediction for capillary-wave turbulence involving four-wave interactions. The carried out high-order correlation analysis (bicoherence and tricoherence) confirms quantitatively the dominant role of four-wave quasi-resonant interactions. The Kolmogorov-Zakharov spectrum constant is also estimated numerically. We interpret our results as the first numerical observation of anisotropic capillary-wave turbulence in which four-wave interactions play a dominant role.

\end{abstract}

\vspace{2mm}

\twocolumn

\normalsize

\textbf{Introduction.}
Wave turbulence occurs when nonlinear waves interact with each other. This phenomenon is ubiquitous: oceanographic waves, plasma waves in the solar wind, nonlinear optical waves, elastic waves, or gravitational waves \cite{ZakharovBook1992,naz2011}. A weakly nonlinear theory (or weak turbulence theory) predicts analytically the wave spectra in various domains involving nonlinear waves \cite{ZakharovBook1992,naz2011,Zakh,zakh1968,zakh70}. Such Kolmogorov-Zakharov's (KZ) spectra can be also obtained by dimensional analysis \cite{naz2003}, an approach allowing also the spectrum predictions in hitherto unexplored wave systems, such as electrohydrodynamic \cite{kochurin18,kochurin19}, magnetohydrodynamic \cite{falcon2008,falcon2011,kochurin20}, or hydroelastic wave turbulence \cite{DeikeJFM2013,DeikePRF2017}. Wave turbulence has been mainly studied for isotropic systems, such as gravity-capillary waves on the surface of a fluid (see reviews \cite{FalconDCDSB2010,newell,zakh19,galtier20}). KZ spectrum has been notably confirmed for isotropic capillary wave turbulence both numerically \cite{Push,Push00,korot03,korot04,pan14,falcon14,PanJFM2015} and experimentally \cite{FalconDCDSB2010,mezhov1}. Anisotropic wave turbulence (i.e. when there is a preferred direction of wave propagation) is much less studied. In this case, nonlinear coherent structures can strongly affect the wave spectrum \cite{Push04one,MMT17,MMT15}, whereas the leading order of nonlinear wave interactions can be modified (e.g five-wave interactions in bidirectional gravity waves \cite{dyachenko95}).

Here, we focus on anisotropic capillary-wave turbulence, when the waves propagate along one direction on the fluid surface (quasi-1D geometry). We perform direct numerical simulation of the cubic nonlinear equations of the corresponding hamiltonian. We observe that four-wave interactions play a dominant role in this peculiar geometry, leading to an efficient energy transfer between scales. The power-law wave spectrum found numerically is in good agreement with the prediction that we obtained by dimensional analysis for four-wave interactions. This latter differs from the usual KZ spectrum of isotropic capillary weak turbulence involving three-wave interactions. Note that for capillary waves, additional spatial symmetry imposes four-wave interactions at the leading order \cite{DuringPRL2009}. Pure capillary-wave turbulence can be reached experimentally either in low-gravity environment \cite{FalconEPL2009}, or at the interface of two immiscible fluids of close densities \cite{IssenmannEPL2016,DuringPRL2009}, but has not been performed in a quasi-1D geometry to the best of our knowledge.

\textbf{Theoretical background --} We consider pure capillary waves on the surface of a fluid. The dispersion relation of linear waves then reads
\begin{equation}\label{disp}
\omega(k)=\left(\frac{\sigma}{\rho}\right)^{1/2} k^{3/2},
\end{equation}
with $\omega$ the angular frequency, $k$ the wave number, $\sigma$ the surface tension, and $\rho$ the mass density of the fluid. For weakly nonlinear waves, energy transfers between waves occur due to a $N$-wave resonant interaction process. The dominant nonlinear wave interaction corresponds to the number $N$ defined as the minimal integer for which the $N$-wave resonant conditions are both satisfied
\begin{equation}
\begin{split}
 \omega(k_1)\pm\omega(k_2)\ldots\pm\omega(k_N)&=0 {\rm \ ,} \\
 k_1\pm k_2\ldots\pm k_N&=0 {\rm \ .} \\
 \end{split}\label{resonance}
\end{equation}
For pure capillary waves in 1D, there is no three-wave resonance interaction: only $k_1=k_2=k_3=0$ is solution of (\ref{resonance}) for $N=3$. No nontrivial four-wave resonance exists: $k_1=k_2$ and $k_3=k_4$ is solution for $N=4$ but does not lead to energy exchange between modes in a $2 \leftrightarrow  2$ interaction process. These can be easily proved graphically (e.g. see \cite{naz2011}). Although exact resonance does not exists, a set of quasi-resonances still exists due to the nonlinear spectral widening. Indeed, for finite wave amplitudes, the widening of the dispersion relation authorizes approximated resonance conditions [i.e. (\ref{resonance}) with $\approx 0$ instead of $=0$]. A 1D capillary-wave turbulence regime is thus expected due to quasi-resonances.

Let us now obtain the wave-elevation spectrum prediction of weak turbulence by dimensional analysis following \cite{naz2003}. The wave elevation power spectrum is defined as $S(k)=|\eta_{k}|^2$, where $\eta_{k}$ is the spatial Fourier transform of the wave elevation $\eta(x)$ at the location $x$. For a capillary wave system involving $N$-wave interactions, we assume
\begin{equation}\label{energy}
S(k)=C P^{\frac{1}{N-1}}\left(\frac{\sigma}{\rho}\right)^{X}k^{Y} {\rm \ ,}
\end{equation}
where $C$ is the Kolmogorov-Zakharov constant, $k$ the wave number modulus, and $P$ a constant energy flux to small scales (direct energy cascade). $X$ and $Y$ are obtained dimensionally, and reads
\begin{equation}\label{en_index}
X=\frac{3}{2(1-N)},\quad Y=d-5+\frac{1-2d}{2(N-1)} {\rm \ ,}
\end{equation}
where $d$ is the dimension in the physical space of the energy density $E_k$ ($E_k$ has dimension $L^{6-d}T^{-2}$, $[S(k)]=L^{5-d}$, and $[P]=L^{5-d}T^{-3}$). $S(k)$ is related to the spectral energy density (per unit surface and mass density) by $E(k)=(\sigma/\rho)k^2S(k)$.
Let us now define $D$ as the dimension of Fourier space. In the case of anisotropic geometry, as here, $d$ and $D$ are not equal to each other.
Here we focus on a bidirectional propagation of waves in a quasi-1D geometry, one has thus $D=1$, and $d=2$ because the wave energy is distributed in the 2D space despite of the collinear geometry.

Assuming that the leading order process is three-wave interaction, (\ref{energy})-(\ref{en_index}) with $d=2$, $N=3$ leads to the prediction of the anisotropic ($D=1$) capillary-wave turbulence spectrum
\begin{align}
S(k)&=C^{{\rm 3w}}_{{\rm 1D}} P^{1/2} \left(\frac{\sigma}{\rho}\right)^{-3/4} k^{-15/4}{\rm \ ,}  \label{surf1D3w1}\\
S(\omega)&=\frac{2}{3} C^{{\rm 3w}}_{{\rm 1D}} P^{1/2} \left(\frac{\sigma}{\rho}\right)^{1/6} \omega ^{-17/6} {\rm \ .} \label{surf1D3w2}
\end{align}
Equation (\ref{surf1D3w2}) being obtained from (\ref{surf1D3w1}) with $S(k)dk=S(\omega)d\omega$. Equation (\ref{surf1D3w1}) is similar to the well known Zakharov-Filonenko spectrum of isotropic capillary wave turbulence ($d=D=2$, $N=3$) \cite{Zakh}, since the dimensional analysis is independent of $D$. Equations (\ref{surf1D3w1})-(\ref{surf1D3w2}) are obtained under the assumption of the dominant influence of three-wave interactions.

At the next order (four-wave interactions), the anisotropic ($D=1$) capillary-wave turbulence spectrum reads from (\ref{energy})-(\ref{en_index}) with $d=2$, $N=4$
\begin{align}
S(k)&=C^{{\rm 4w}}_{{\rm 1D}}P^{1/3}\left(\frac{\sigma}{\rho}\right)^{-1/2}k^{-7/2}{\rm \ ,}  \label{surf1d4w1}\\
S(\omega)&=\frac{2}{3} C^{{\rm 4w}}_{{\rm 1D}}P^{1/3}\left(\frac{\sigma}{\rho}\right)^{1/3}\omega^{-8/3} {\rm \ .} \label{surf1d4w2}
\end{align}

Note that the predictions of (\ref{surf1d4w1})-(\ref{surf1d4w2}) involving four-wave interactions differ from the ones of (\ref{surf1D3w1})-(\ref{surf1D3w2}) obtained for $N=3$. The difference in the $k$-power-law exponent is roughly 7\%: $-15/4=-3.75$ ($N=3$) vs. $-7/2=-3.5$ ($N=4$). Since (\ref{en_index}) does not depend on $D$, (\ref{surf1d4w2}) is in agreement with the solution derived theoretically for four-wave interactions of isotropic capillary waves \cite{DuringPRL2009}. The aim of this work is to perform direct numerical simulation of anisotropic capillary-wave turbulence with high accuracy. We show that a capillary-wave turbulence regime is observed in this peculiar 1D bidirectional geometry, once a high enough level of pumping is reached. We show that the $k$-scaling of the wave elevation spectrum is very close to the prediction of (\ref{surf1d4w1}) for the four-wave interaction case.

\textbf{Model equations --} Consider the irrotational potential flow of an ideal incompressible fluid of infinite depth. The numerical model used here is based on the well-known hamiltonian equations of motion of an ideal fluid with the free surface  obtained by Zakharov et al. \cite{Zakh,zakh1968}. In this approach, the original three-dimensional equations of hydrodynamics are transformed to weakly nonlinear equations that directly describe the dynamics of the free surface in two spatial coordinates $x$ and $y$. Here, we neglect gravity and only the dynamics of capillary waves is considered.

The Hamiltonian of the system under study in dimensionless units ($\rho=1$, $\sigma=1$) reads
\begin{equation}
\begin{split}
H&=\frac{1}{2}\int \left(\sqrt{1+(\nabla \eta)^2}-1\right)d\textbf{r}\\
&+\frac{1}{2}\int \left(\psi \hat k \psi+\eta \left[|\nabla \psi|^2-(\hat k \psi)^2\right]\right)d\textbf{r} \\
&+\frac{1}{2} \int \eta (\hat k \psi)\left[\hat k(\eta \hat k \psi)+\eta\nabla^2 \psi\right]d\textbf{r} \ {\rm \ ,}\\
\end{split}\label{Hal}
\end{equation}
with $d\textbf{r}=dxdy$. The function $\psi(x,y,t)$ is the value of the velocity potential on the boundary of the liquid,  $\nabla=(\partial_x,\partial_y)$ is the nabla operator, and $\hat k $ is the integral operator defined in Fourier space as $\hat k f_\textbf{k}=|\textbf{k}| f_\textbf{k}$. The equations of the boundary motion are obtained by taking the variational derivatives:
\begin{equation}\label{eq}\frac{\partial \eta}{\partial t}=\frac{\delta H}{\delta \psi},\qquad\frac{\partial \psi}{\partial t}=-\frac{\delta H}{\delta \eta}.
\end{equation}
The equations (\ref{eq}) of the boundary motion  are applicable for any symmetry of the surface perturbations. In the quasi-1D geometry they can be simplified. Since, the functions $\eta_\textbf{k}$ and $\psi_\textbf{k}$ depend only on $x$-component of the wave vector, the absolute value of $\textbf{k}$ is given by $k=|\textbf{k}|=|k_x|$, and the nabla operator is reduced to the partial derivative $\nabla\to\partial_x$. In such geometry, we can integrate the right-hand side in (\ref{Hal}) with respect to $y$ and introduce the normalized hamiltonian as, $H_{1D}=H/L_y$, where $L_y$ is the characteristic length of the system in the direction normal to the wave propagation.

For a complete description of the turbulent motions, we add to the equations (\ref{eq})  the terms responsible for the pumping and dissipation of energy. As a result, the cubically nonlinear equations (\ref{eq}) of the boundary motions take the form
\begin{multline}
\eta_t=\hat k \psi-(\eta \psi_x)_x-\hat k(\eta \hat k \psi)+\hat k(\eta \hat k[\eta \hat k \psi])\\
           +\frac{1}{2} (\eta^2 \hat k \psi)_{xx}+\frac{1}{2}\hat k (\eta^2  \psi_{xx})+\hat D_k\eta  \ {\rm \ ,} \label{eta1D}
\end{multline}
\begin{multline}
\psi_t=\eta_{xx}-\frac{1}{2}\left[(\psi_x)^2-(\hat k \psi)^2\right] \\
           -(\hat k \psi)\hat k(\eta \hat k \psi)-(\eta \hat k \psi)\psi_{xx}+F(x,t)+\hat D_k \psi  {\rm \ .}   \label{psi1D}
\end{multline}
The operator $\hat D_k$ is responsible for the effect of viscosity, in $k$-space it is defined as,
\[\hat D_k=-\nu_0(|k|-k_d)^2,\quad |k|\geq k_d;\quad \hat D_k=0, \quad |k|\leq k_d.\]
Here $\nu_0$ is a constant, and $k_d$ is the wave number determining the spatial scale at which the energy dissipation occurs. The term responsible for the pumping of energy is defined as follows,
\begin{equation}\label{pump}F(x,t)=a_0\sum_{i=1}^{4} \sin(f_it)\cos(k_ix+\phi_i),\end{equation}
where $a_0$ is a constant defining the forcing strength, $k_i=i$ are the forcing wave numbers, $f_i$ and $\phi_i$ are the random numbers normally distributed in the interval $[0,2\pi]$. The forcing term thus corresponds to a superimposition of few oscillating stationary waves each of random amplitudes and phases in the following range of the forcing scale: $k_i=1$, 2, 3 and 4. To stabilize the numerical scheme, we retain only the linear term in the expansion of the boundary curvature in (\ref{psi1D}), see \cite{korot} for more details.

\begin{figure}[t]
\centering
\includegraphics[width=0.9\linewidth]{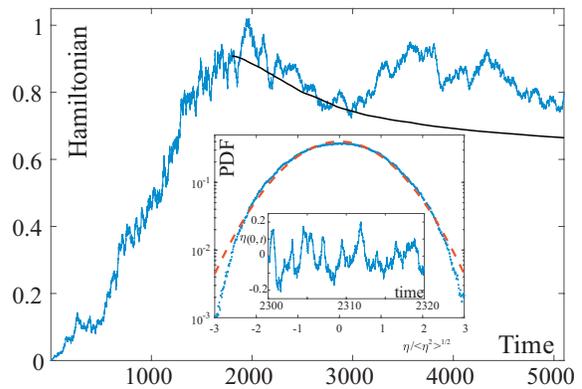}
\caption{(Color online) Normalized hamiltonian $H_{1D}$ vs time. Blue solid line corresponds to forcing amplitude $a_0=2$, and black line to a decaying simulation (no forcing $a_0=0$). Inset: PDF of wave height rescaled by its standard deviation (blue dots), red dashed line corresponds to Gaussian fit; the inset shows the temporal evolution of the wave height $\eta(t)$ at $x=0$.}
\label{fig1}
\end{figure}

\textbf{Simulation results --} The numerical integration scheme for the solution of the system of (\ref{eta1D})--(\ref{psi1D}) in time is based on the fourth-order explicit Runge-Kutta method with the step $dt$. The spatial derivatives and integral operators are calculated using the pseudo-spectral methods with the total number harmonics $\mathcal{N}$. The simulations were performed in a periodic 1D box of size $L_x=2\pi$ with the parameters: $x\in [0,L_x]$, $\mathcal{N}=2048$, $dt=2.5\ 10^{-5}$, $a_0=2$, $k_i=1$, 2, 3, and 4, $k_d=500$, and $\nu_0=10$.

We present first numerical simulations corresponding to a continuous energy pumping over time. Figure 1 shows the temporal evolution of the normalized hamiltonian $H_{1D}$ (the total energy of the system). It first increases with time and then reaches a quasi-stationary state for $t\geq 2000$ (see blue solid line). Within this quasi-stationary state, the wave height $\eta(t)$ at fixed location is found to be erratic (see inset of Fig.~\ref{fig1}), and its probability density function (PDF) computed over $t\in[2000, 5000]$ is close to Gaussian distribution (see inset of Fig.~\ref{fig1}). The time-averaged value of the wave steepness in the quasi-stationary state is $\left\langle \sqrt{\int_{L_x}|\partial \eta(x,t)/ \partial x|^2dx/L_x}\right\rangle \simeq 0.36$, $\langle \cdot \rangle$ denoting a temporal averaging. Although the latter is not weak, the distribution is found to be symmetric (Skewness $Sn\equiv \langle \eta^3\rangle/\langle \eta^2\rangle^{3/2}\simeq0.01 $), whereas the extreme events are less likely than for a Gaussian distribution (Kurtosis $\langle \eta^4 \rangle/\langle \eta^2 \rangle^{2}\simeq 2.65<3$) underlying an unexpected symmetric light-tailed distribution. Indeed, contrary to gravity or gravity-capillary waves where nonlinearities sharpen the crests and flatten the troughs (i.e. $Sn> 0$), highly symmetric distribution is observed here (as for isotropic capillary-wave turbulence for 3-wave \cite{falcon14} or 4-wave interactions \cite{DuringPRL2009} for comparable wave steepness, but with light tails due to nonlinear effects.

\begin{figure}[t]
\centering
\includegraphics[width=0.9\linewidth]{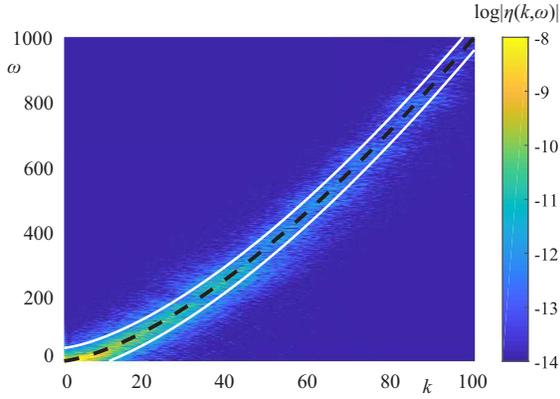}
\caption{(Color online) Space-time Fourier power spectrum $S(k,\omega)$ of wave elevation. Black dashed line corresponds to the linear dispersion relation of (\ref{disp}), and white solid lines show a nonlinear frequency broadening: $\omega(k)\pm \delta_\omega$ with $\delta_\omega=40$.}
\label{fig2}
\end{figure}

\emph{Dispersion relation --} Figure \ref{fig2} shows the full space-time Fourier power spectrum $S(k,\omega)$ of the wave elevation $\eta(x,t)$ in the stationary state ($t\in[2000,5000]$). We observe that the most of the energy injected at low wave numbers is transferred within a large range of wave numbers, and is concentrated around the linear dispersion relation (\ref{disp}) as expected for weak capillary-wave turbulence. Energy is also present out of this law corresponding to a spectral widening due to nonlinear effects. Since no wave resonant interaction is authorized in this 1D system (see above), this broadening is consistent with an energy transfer due to quasi-resonance (see below). No signature of coherent structures appears on this nonlinear dispersion relation such as bound waves as reported routinely in isotropic gravity or gravity-capillary wave turbulence \cite{HerbertPRL10,BerhanuPRE13,CampagnePRF19}. No deviation from (\ref{disp}) is also observed due to nonlinear corrections \cite{CrapperJFM57,BerhanuPRE13}.

\begin{figure}[t]
\centering
\includegraphics[width=0.9\linewidth]{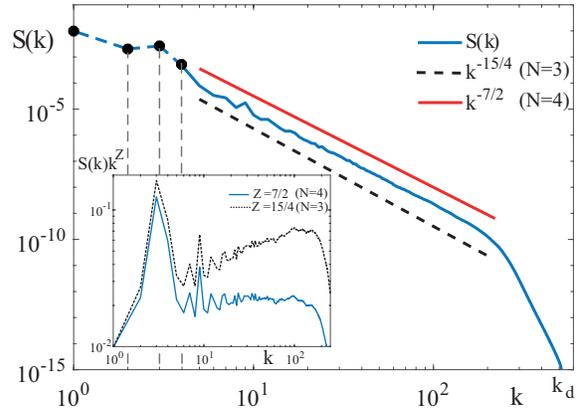}
\caption{(Color online)  Time-averaged spatial spectrum $S(k)$ of wave elevation in the quasi-stationary state. Black bullets correspond to the harmonics pumped. Solid line corresponds to (\ref{surf1d4w1}), dashed line to (\ref{surf1D3w1}). Inset: compensated spectra $S(k)k^{15/4}$ and $S(k)k^{7/2}$ vs. $k$.}
\label{fig3}
\end{figure}

\emph{Spatial spectrum --} Figure \ref{fig3} shows the time-averaged spatial-power spectrum $S(k)$ of the wave height $\eta(x,t)$ in the stationary state. A clear power-law scaling is observed on more than one decade in $k$. The best fit is $S(k)\sim k^{-3.5\pm 0.1}$ which is closer to the prediction of (\ref{surf1d4w1}) than to the one of (\ref{surf1D3w1}). The compensated spectra are shown in the inset of Fig. \ref{fig3}. Indeed, the wave turbulence spectrum $S(k)$ is better approximated by $k^{-7/2}$ ($N=4$) than by $k^{-15/4}$ ($N=3$) within the inertial range $5<k<180$. Thus, our direct numerical simulations suggest that four-wave quasi-resonant interactions are involved for 1D capillary-wave turbulence. Let us prove it below indirectly.

\emph{Quasi-resonances --} High-order correlation analysis is now performed to confirm quantitatively the influence of four-wave quasi-resonant interactions in a consistent way with the above results from the wave spectrum (second-order correlation). As explained above, the three-wave resonance conditions (\ref{resonance}) are not satisfied exactly in 1D, but the occurrence of quasi-resonances due to nonlinear spectral widening is still possible. Write the three-wave quasi-resonance condition in $\omega$,
\begin{equation}
|(k_1+k_2)^{3/2}-k_1^{3/2}-k_2^{3/2}|\leq \delta_{\omega} {\rm \ ,}
\label{3wave}
\end{equation}
where $\delta_{\omega}$ is the nonlinear wave frequency shift, usually estimated by nonlinear spectral widening as in Fig.~\ref{fig2}. The condition for the four-wave quasi-resonance reads
\begin{equation}
|(k_1+k_2-k_3)^{3/2}-k_1^{3/2}-k_2^{3/2}+k_3^{3/2}|\leq\delta_{\omega} {\rm \ .}
\label{4wave}
\end{equation}
To check the existence of 3-wave quasi-resonant interactions satisfying (\ref{3wave}), we compute the normalized third-order correlation in $k$ of the wave height (the so-called bicoherence) \cite{PunzmannPRL2009,aub16}
\begin{equation}
B(k_1,k_2)=\frac{|\langle \eta^*_{k_1}\eta_{k_2}\eta_{k_1+k_2}\rangle|}{\sqrt{\langle|\eta_{k_1}|^2\rangle\langle|\eta_{k_2}\eta_{k_1+k_2}|^2\rangle}}  {\rm \ ,}
\label{bic}
\end{equation}
where $\ast$ denotes the complex conjugate. To highlight 4-wave quasi-resonant interactions satisfying (\ref{4wave}), we compute the fourth-order correlation of the wave height (or tricoherence) \cite{aub17}
\begin{equation}
T(k_1,k_2,k_3)=\frac{|\langle\eta^*_{k_1}\eta^*_{k_2}\eta_{k_3}\eta_{k_1+k_2-k_3}\rangle|}{\sqrt{\langle|\eta_{k_1}\eta_{k_2}|^2\rangle\langle|\eta_{k_3}\eta_{k_1+k_2-k_3}|^2\rangle}} {\rm \ .}
\label{tric}
\end{equation}

\begin{figure}[t]
\centering
\includegraphics[width=1.0\linewidth]{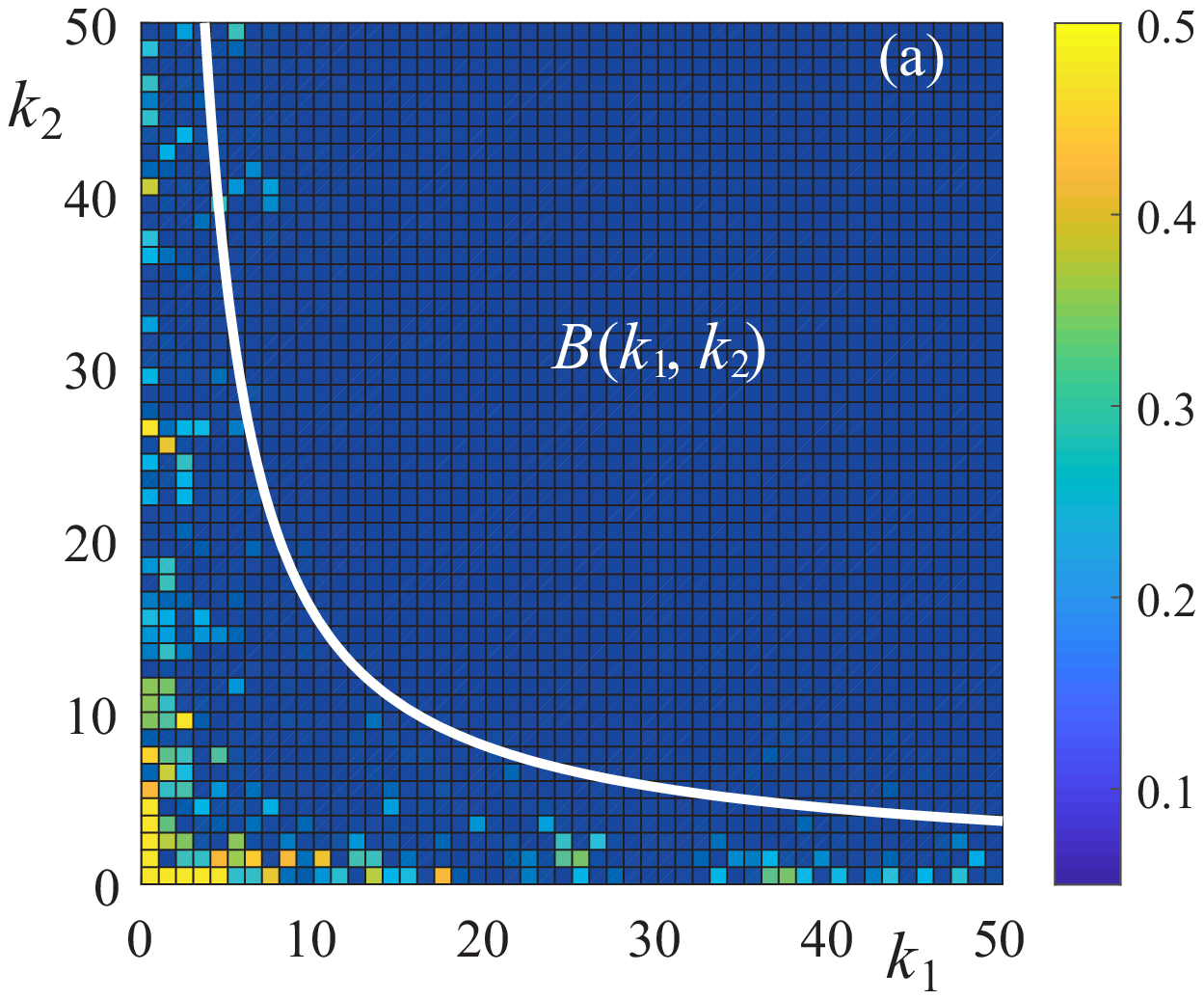}
\includegraphics[width=1.0\linewidth]{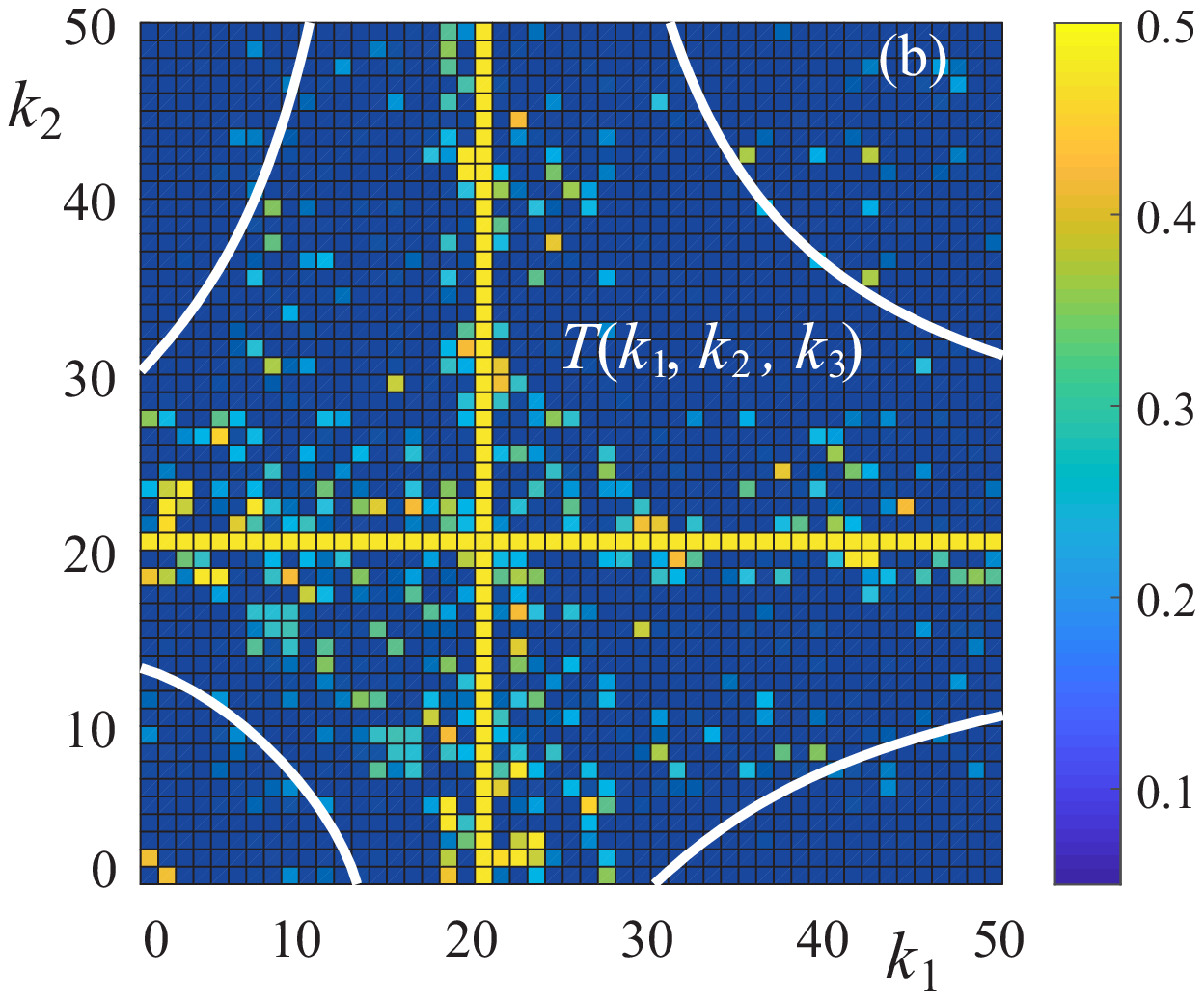}
\caption{(Color online) Bicoherence (a) and tricoherence (b) of the Fourier transform of the wave elevation estimated from (\ref{bic}) and (\ref{tric}), resp. The tricoherence is shown for the fixed value $k_3=20$. White lines correspond to (\ref{3wave}) in (a) [resp. (\ref{4wave}) in (b)] with $\delta_{\omega}=40$, i.e. to the boundaries of authorized 3-wave [resp. 4-wave] quasi-resonances.}
\label{fig4}
\end{figure}

Due to the peculiar normalizations used in (\ref{bic})-(\ref{tric}), the functions $B(k_1, k_2)$ and $T(k_1,k_2,k_3)$ range from 0 (no correlation) to 1 (perfect correlation). $B(k_1, k_2)$ and $T(k_1,k_2,k_3)$ are computed in the quasi-stationary regime, and are displayed in Fig. \ref{fig4}. Figure \ref{fig4}(a) shows that three-wave quasi-resonances are indeed present but their number is very small: most of the quasi-resonant modes are located at low wave numbers near forcing ones within the area bounded by (\ref{3wave}) (white line). Figure \ref{fig4}(b) shows a completely different situation. We can see a large number of quasi-resonant modes within the area bounded by (\ref{4wave}) (white lines) as well as degenerated 4-wave exact resonances at $k_1=k_3$, and $k_2=k_3$.  Quantitatively, the ratio between the sum of tricoherence values (at fixed $k_3$) excluding trivial resonances and the bicoherence ones is $r=\int \int \frac{T(k_1,k_2,20)}{B(k_1,k_2)}dk_1dk_2\simeq 1.6$. Taking into account all possible values of $k_3$ in the inertial range leads to $r\simeq70$. Thus, the energy transfers are mostly due to four-wave quasi-resonant interactions.

\begin{figure}[t]
\centering
\includegraphics[width=0.9\linewidth]{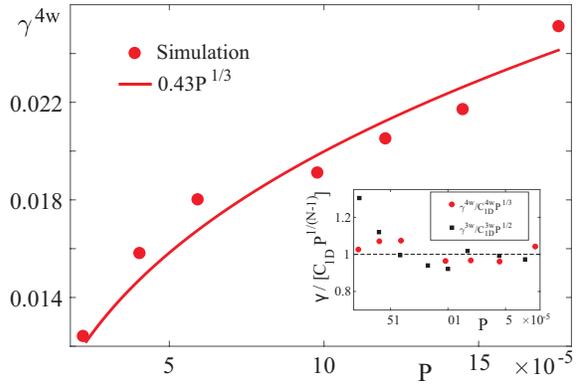}
\caption{(Color online) Spectrum coefficient $\gamma^{{\rm 4w}}(P)$ versus energy dissipation rate $P$ (red markers). Solid line corresponds to the fit $0.43 P^{1/3}$. Inset: compensated Kolmogorov-Zakharov's constants versus $P$ for a 3-wave (black square) or 4-wave (red circle) process.}
\label{fig5}
\end{figure}

\emph{Kolmogorov-Zakharov's constant --} One can finally estimate the spectrum constant $C^{{\rm 4w}}_{{\rm 1D}}$ of (\ref{surf1d4w1}) from simulation series in decaying wave turbulence. To wit, after $t=1800$ from the beginning of the simulations, we suddenly stop the pumping. The system then slowly loses energy from this wave turbulence regime (see black line in Fig. \ref{fig1}). Rewriting the weak turbulence spectrum prediction of (\ref{surf1d4w1}) as $S(k)=\gamma^{{\rm 4w}} k^{-7/2}$ where the coefficient $\gamma^{{\rm 4w}}$ depends on the energy dissipation rate $P$ and on the constant $C^{{\rm 4w}}_{{\rm 1D}}$ as $\gamma^{{\rm 4w}}(P)=C^{{\rm 4w}}_{{\rm 1D}}P^{1/3}$ (since $\sigma=\rho=1$). Figure~\ref{fig5} shows the numerically obtained dependence of $\gamma^{{\rm 4w}}$ with $P$ during the decay ($P$ being estimated from the solid black line in Fig. \ref{fig1}, and $\gamma^{{\rm 4w}}$ from the decaying spectrum). By fitting this coefficient with a $P^{1/3}$ dependence, we infer Kolmogorov-Zakharov's constant $C^{{\rm 4w}}_{{\rm 1D}}$ as $0.43\pm 0.04$. This value could be compared to the theoretical constant that can, in principle, be computed \cite{DuringPRL2009}. Although 3-wave quasi-resonance interactions are not the dominant process, we perform an analogous analysis for a 3-wave process leading to the numerical value for $C^{{\rm 3w}}_{{\rm 1D}}=5.9\pm 0.06$. The inset of Fig.~\ref{fig5} shows the compensated values of the Kolmogorov-Zakharov's constants for the two interaction processes.

\emph{Scale separation --} Weak turbulence theory assumes a timescale separation between the linear timescale $\tau_{{\rm lin}}=1/\omega$, and the nonlinear time, which reads dimensionally $\tau_{{\rm nl}}^{{\rm 4w}}\sim P^{-2/3}(\sigma/\rho)^{1/2}k^{-1/2}$ for capillary waves ($N=4$, $d=2$, $\forall D$). This means a nonlinearity parameter $\tau_{{\rm lin}}/\tau_{{\rm nl}}^{{\rm 4w}} \sim P^{2/3}(\sigma/\rho)^{-1}k^{-1} \ll 1$. As nonlinearity increases with $P$, breaking of weak turbulence is thus expected to occur for $P > P_c=k^{3/2}(\sigma/\rho)^{3/2}$ ($P_c$ is independent on $N$). The maximum value of $P$ used here is few orders of magnitude less than this critical flux regardless the value of $k$, and thus satisfies indirectly the expected scale separation.

\emph{Conclusion --} Although resonant wave interactions are forbidden in this geometry, we show that anisotropic (quasi-1D) pure capillary waves display small-scale wave turbulence due to four-wave quasi-resonant interactions, as highlighted by the wave spectrum scaling and high-order correlations analysis. An unexpected symmetric light-tailed wave distribution is also observed. We hope that our study will trigger future investigations, notably to better understand the large-scale dynamics (larger than the forcing scale) of collinear wave turbulence such as the inverse cascade or the statistical equilibrium.

\textbf{Acknowledgments --}  E.K. thanks partial support by Russian Science Foundation project No.~19-71-00003. E.F. thanks partial support of the French National Research Agency (ANR Dysturb, project No. ANR-17-CE30-0004), and of the Simons Foundation/MPS No. 651463-Wave Turbulence notably for the mission of E.K. in Paris, France. Software tool development for numerical simulation was partially supported by RFBR, project No.~20-38-70022.

\end{document}